\documentclass[prd,aps,onecolumn,superscriptaddress,nofootinbib,showpacs,a4paper,10pt, notitlepage]{revtex4-1}

\usepackage{amsmath,amscd,amssymb,amsfonts}
\newcommand\id{\leavevmode\hbox{\small1\kern-3.3pt\normalsize1}}
\usepackage{graphicx}
\usepackage[labelformat=default, labelsep=default, justification=centerlast,  font=small]{caption}
\usepackage{sidecap}
\usepackage{epstopdf}
\usepackage{geometry}
\usepackage{float}
\usepackage{verbatim}
\usepackage{psfrag}
\usepackage{epsfig}
\usepackage{cancel}
\usepackage{textcase}
\usepackage{pstricks}
\usepackage{xcolor}

\newcommand{\clock}{\mbox{
\begin{picture}(0,5)
\put(.5,0){\circle{7}}
\put(0.1,0){\line(0,1){2.4}}
\put(0,0){\line(1,0){1.6}}
\end{picture}}}

\newcommand{\h}{H_{\!\clock\;}\;}
\newcommand{\hlab}{H_{\mbox{\scriptsize Lab}}}

\newcommand{\bra}{\langle}
\newcommand{\ket}{\rangle}

\def\ms{m$\cdot$s }

\begin {document}

\title{Quantum interferometric visibility as a witness of  general relativistic proper time}

\author{Magdalena Zych*}
\email[Magdalena Zych: ]{magdalena.zych@univie.ac.at}
\author{Fabio Costa}
\email[Fabio Costa: ]{fabio.costa@univie.ac.at}
\author{Igor Pikovski}
\email[Igor Pikovski: ]{igor.pikovski@univie.ac.at}
\affiliation{Faculty of Physics, University of Vienna, Boltzmanngasse 5, 1090 Vienna, Austria}
\author{\v{C}aslav Brukner}
\email[\v{C}aslav Brukner: ]{caslav.brukner@univie.ac.at}
\affiliation{Faculty of Physics, University of Vienna, Boltzmanngasse 5, 1090 Vienna, Austria}
\affiliation{Institute for Quantum Optics and Quantum Information, Austrian Academy of Sciences, Boltzmanngasse 3,  1090 Vienna, Austria}


\begin{abstract}
Current attempts to probe general relativistic effects in quantum mechanics focus on precision measurements of phase shifts in matter-wave interferometry. Yet, phase shifts can always be explained as arising due to an Aharonov-Bohm effect, where a particle in a flat space-time is subject to an effective potential. Here we propose a novel quantum effect that cannot be explained without the general relativistic notion of proper time.  We consider interference of a "clock" - a particle with evolving internal degrees of freedom - that will not only display a phase shift, but also reduce the visibility of the interference pattern. According to general relativity proper time flows at different rates in different regions of space-time. Therefore, due to quantum complementarity the visibility will drop to the extent to which the path information becomes available from reading out the proper time from the "clock". Such a gravitationally induced decoherence would provide the first test of the genuine general relativistic notion of proper time in quantum mechanics.
\end{abstract}
\maketitle

\section*{Introduction}\label{sec:intro}
In the theory of general relativity time is not a global background parameter, but flows at different rates depending on the space-time geometry.  Although verified to high precision in various experiments \cite{hafele}, this prediction (as well as any other general relativistic effect) has never been tested in the regime where quantum effects become relevant. There is, in general, a fundamental interest in probing the interplay between gravity and quantum mechanics \cite{chiao}. The reason is that the two theories are grounded on seemingly different premises and, although consistent predictions can be extrapolated for a large range of phenomena, a unified framework is still missing and fundamentally new physics is expected to appear at some scale.

\begin{figure}[ht]
 \vspace{-5pt}
\begin{center}
\includegraphics[width=8.5cm]{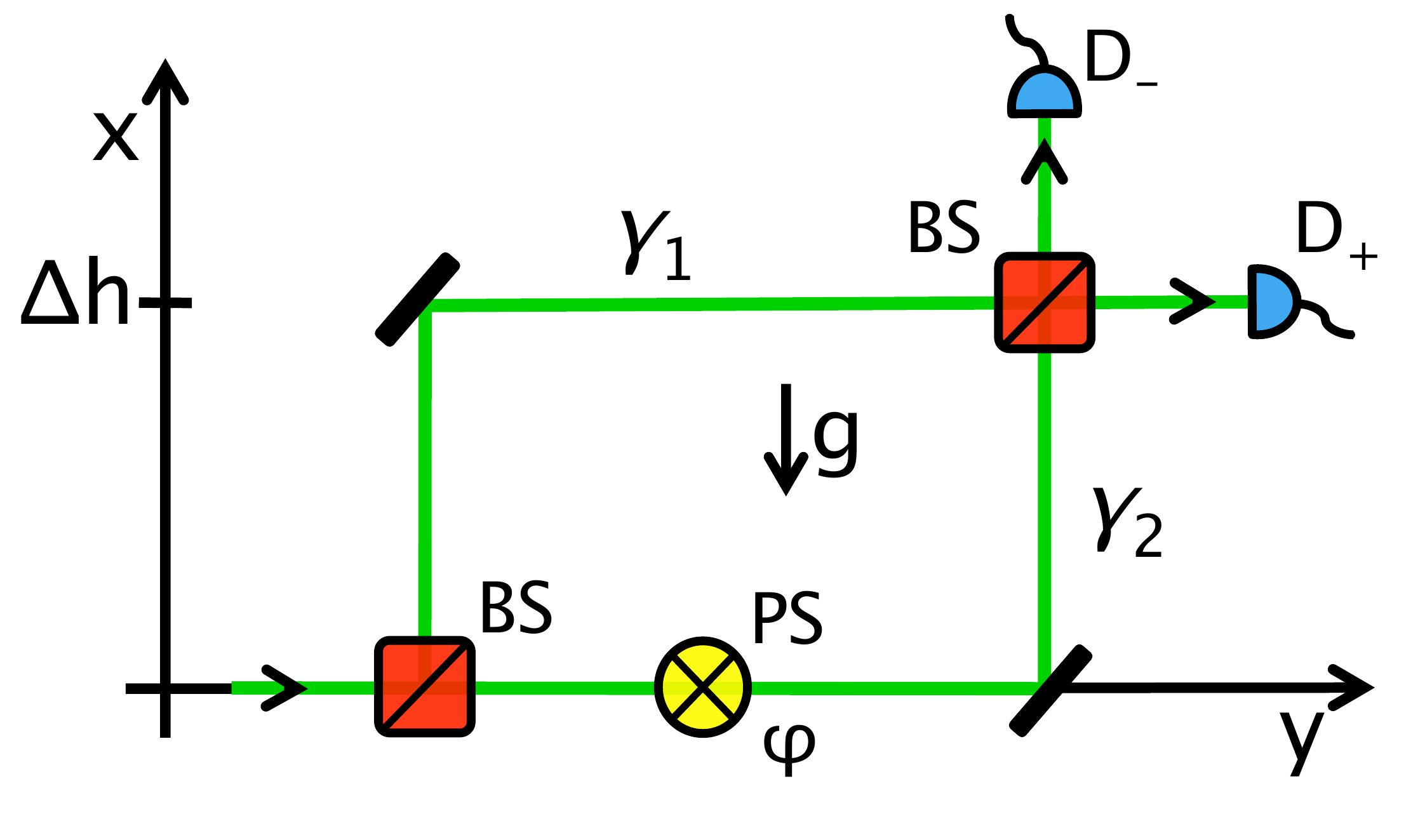}
\end{center}
\vspace{-25pt}
\caption{The Mach-Zehnder interferometer in a gravitational field. The setup  considered in this work consists of two beam splitters (BS), a phase shifter (PS) and two detectors $D_\pm$.  The PS gives a controllable phase difference $ \varphi$ between the two trajectories $\gamma_1$ and $\gamma_2$, which both lie in the $x-y$ plane. A homogeneous gravitational field ($g$) is oriented antiparallel to the $x$ direction. The separation between the paths in the direction of the field is $\Delta h$.  General relativity predicts that the amount of the elapsed proper time is different along the two paths.  In our approach we will consider interference of a particle (which is not in a free fall) that has an evolving internal degree of freedom which acts as a ``clock''.  Such an interference experiment will therefore not only display a phase shift, but also reduce the visibility of the interference pattern to the extent to which the path information becomes available from reading out the proper time of the ``clock''.\hspace*{\fill}}
\label{machzehnder}
\end{figure}

One of the promising experimental directions is to reveal, through interferometric measurements, the phase acquired by a particle moving in a gravitational potential \cite{wajima, dimopoulos}. Typically considered is a  Mach-Zehnder type interferometer, see Fig.\ \ref{machzehnder}, placed in the Earth's gravitational field, where a particle  travels in a coherent superposition along the two interferometric paths $\gamma_1$, $\gamma_2$ which have different proper length. The two amplitudes in the superposition acquire different, trajectory dependent phases $\Phi_i$, $i=1,2$. In addition, the particle acquires a controllable relative phase shift $\varphi$. Taking into account the action of the first beam splitter and denoting by $|r_i\ket$ the mode associated with the respective path $\gamma_i$,  the state inside the Mach-Zehnder setup $|\Psi_{MZ}\ket$, just before it is recombined, can be written as
\begin{equation}
\label{intermed}
|\Psi_{MZ}\ket=\frac{1}{\sqrt{2}}\left(i e^{-i\Phi_1}|r_1\ket+ e^{-i\Phi_2+i\varphi}|r_2\ket\right).
\end{equation}
Finally, the particle can be registered by one of the two detectors  $D_\pm$ with corresponding probabilities  $P_\pm$:
\begin{equation}
\label{probab_intro}
P_\pm=\frac{1}{2}\pm \frac{1}{2}\cos\left(\Delta\Phi+\varphi\right),
\end{equation}
where $\Delta\Phi:=\Phi_1-\Phi_2$. The phase $\Phi_i$ is proportional to the action along the corresponding (semiclassical) trajectory $\gamma_i$ on which the particle moves. For a free particle on an arbitrary space-time background the action can be written in terms of the proper time $\tau$ that elapsed during the travel, $S_i=-mc^2 \int_{\gamma_i}d\tau$. This might suggest that the measurement of $\Delta \Phi$ is an experimental demonstration of the general relativistic time dilation.

There is, however, a conceptual issue in interpreting experiments measuring a gravitationally induced phase shift as  tests of the relativistic time dilation. The action $S_i$ above can be written in terms of an effective gravitational potential on a flat space-time. Thus, all the effects resulting from such an action are fully described by the Sch\"odinger equation with the corresponding  gravitational potential and where the time evolution is given with respect to the global time.  Note that a particle in a field of arbitrary nature is subject to a Hamiltonian where the potential energy is proportional to the field's charge and a position dependent potential. Therefore, even in a homogeneous field the particle acquires a trajectory dependent phase although the force acting on it is the same at any point - the phase arises only due to the potential. For a homogeneous electric field this relative phase is known as the electric Aharonov-Bohm effect \cite{aharonov}. The case of Newtonian gravity is directly analogous - the role of the particle's electric charge and of the Coulomb potential are taken by the particle's mass and the Newtonian gravitational potential, respectively \cite{morgan}. All quantum interferometric experiments performed to date (see e.g.\ Refs.\ \cite{cow, rauch, chu}) are fully explainable by this gravitational analogue of the  electric Aharonov-Bohm effect.  Moreover, even if one includes non-Newtonian terms in the Hamiltonian, this dichotomy of interpretations is still present. Again, one can interpret the phase shift $\Delta\Phi$ as a type of  an Aharanov-Bohm phase, which a particle moving in a \emph{flat} space-time acquires due to an effective, non-Newtonian, gravitational potential (at least for an effective gravitational potential arising from the typically considered Kerr or Schwarzschild space-times).


Here we predict a quantum effect that cannot be explained without the general relativistic notion of proper time and thus show how it is possible to unambiguously distinguish between the two interpretations discussed above.  We consider a Mach-Zehnder interferometer placed in the gravitational potential  and with a ``clock'' used as an interfering particle.  By ``clock'' we mean some evolving internal degree of freedom of the particle. If there is a difference in proper time elapsed along the two trajectories, the ``clock'' will evolve into different quantum  states for the two paths of the interferometer. Due to quantum complementarity between interference and which-path information the interferometric visibility will decrease by an amount given by the which-way information accessible from the final state of the clock \cite{englert, green_yasin, wooters}.  Such a reduction in the visibility is a direct consequence of the general relativistic time dilation, which follows from the Einstein equivalence principle. Seeing the Einstein equivalence principle as a corner stone of general relativity, observation of the predicted loss of the interference contrast would be the first confirmation of a genuine general relativistic effect in quantum mechanics.

One might sustain the view that the interference  observed with particles without evolving degrees of freedom is a manifestation of some intrinsic oscillations associated with the particle and that such oscillations can still be seen as the ticking of a clock which keeps track of the particle's time. If any operational meaning was to be attributed to this clock, it  would  imply that  which-way information is in principle accessible.  One should then either assume that proper time is a quantum degree of freedom, in which case there should be a drop in the interferometric visibility, or that the quantum complementarity relation (between which-path information and interferometric visibility) would be violated when general relativistic effects become relevant. Our proposed experiment allows to test these possibilities. The hypothesis that proper time is a degree of freedom has indeed been considered in various works \cite{greenberger1, greenberger2, matsumoto}.  

The above considerations are also relevant in the context of the debate over Ref.\ \cite{muller} (determination of the gravitational redshift by reinterpreting interferometric experiment \cite{chu} that measured the acceleration of free fall). It was pointed out in Refs.\ \cite{wolf_rep, sinha, giulini, wolf}  that only states non-trivially evolving in time can be referred to as ``clocks''.  In Ref.\ \cite{sinha} the interference in such a case was discussed, however, the role of the interferometric visibility as a witness of proper time in quantum mechanics and as a tool to test new hypotheses has not been previously considered. 

In the present paper we discuss an interferometric experiment in the gravitational field where the interfering particle can be operationally treated as a ``clock''.  We predict that
as a result of the quantum complementarity between interference and which-path information the general relativistic time dilation will cause the decrease in the interferometric visibility. The observation of such a reduction in the visibility  would be the first  confirmation of a genuinely general relativistic effect in quantum mechanics, in particular it would unambiguously probe the proper time as predicted by general relativity. The proposed experiment can also lead to a conclusive test of theories in which proper time is treated as a quantum degree of freedom. 
 

\section*{Results}\label{sec:results}
\subsection*{Which-way information from proper time.}\label{sec:main}
Consider an interferometric experiment with the setup as in Fig.\ \ref{machzehnder}, but in a situation where the particle in superposition has some internal degree of freedom which can evolve in time. In such a case state \eqref{intermed} is no longer the full description of the system. Moreover, if this degree of freedom can be considered as  a ``clock'', according to the general relativistic notion of proper time it should evolve differently along the two arms of the interferometer in the presence of gravity. For a trajectory $\gamma_i$ let us call $|\tau_i\ket$ the corresponding state of the ``clock''. The superposition  \eqref{intermed} inside the interferometer now reads
\begin{equation}
\label{intermed_tau}
|\Psi_{MZ}\ket=\frac{1}{\sqrt{2}}\left(ie^{-i\Phi_1}|r_1\ket |\tau_1\ket +e^{-i\Phi_2+i\varphi}|r_2\ket |\tau_2\ket \right).
\end{equation}
In general, the state \eqref{intermed_tau} is entangled and according to quantum mechanics interference in the path degrees of freedom should correspondingly be washed away. The reason is that one could measure the ``clock''  degrees of freedom and in that way read out the accessible which-path information. Tracing out  the ``clock'' states in Eq.\ \eqref{intermed_tau} gives the detection probabilities 
\begin{equation}
\label{probab_intro_tau}
P_{\pm}=\frac{1}{2}\pm \frac{1}{2}|\bra\tau_1|\tau_2\ket|\cos\left(\Delta\Phi+\alpha + \varphi \right),
\end{equation}
where $\bra\tau_1|\tau_2\ket = |\bra\tau_1|\tau_2\ket|e^{i\alpha}$.  When the ancillary phase shift $\varphi$ is varied, the probabilities $P_\pm$ oscillate with the amplitude $\mathcal V$, called the visibility (contrast) of the interference pattern. Formally $\mathcal V:=\frac{Max_{\varphi} P_\pm - Min_{\varphi} P_\pm}{Max_{\varphi} P_\pm + Min_{\varphi} P_\pm}$.  Whereas without the ``clock'' the expected contrast is always maximal (Eq.\ \eqref{probab_intro} yields $\mathcal V=1$), in the case of Eq.\ \eqref{probab_intro_tau} it reads
\begin{equation}
\label{visib_tau}
\mathcal V=|\bra\tau_1|\tau_2\ket|.
\end{equation}
The distinguishability $\mathcal D$ of the trajectories is the probability to correctly guess which path was taken in the two-way interferometer by measuring the degrees of freedom that serve as a which-way detector  \cite{englert} (in  mathematical terms it is the trace norm distance between the final states of the detectors associated with different paths). In our case these are the ``clock'' degrees of freedom and we obtain $\mathcal D=\sqrt{ 1-|\bra \tau_1|\tau_2\ket|^2}$. The amount of the which-way information that is potentially available sets an absolute upper bound on the fringe visibility and we recover the  well known duality relation \cite{englert, green_yasin, wooters} in the form $\mathcal V^2+\mathcal D^2=1$, as expected for pure states. 
 
The above result demonstrates that general relativistic effects in quantum interferometric experiments can go beyond previously predicted corrections to the non-relativistic phase shift. When proper time is treated operationally we anticipate the gravitational time dilation to result in the reduction of the fringe contrast. This drop in the visibility is expected independently of how the proper time is measured and which system and interaction are used for the ``clock''.  
Moreover when the information about the time elapsed is not physically accessible the drop in the visibility will not occur. This indicates that the effect unambiguously arises due to the proper time as predicted by general relativity, in  contrast to measurements of the phase shift alone.  The gravitational phase shift  occurs independently of whether the system can or cannot be operationally treated as a ``clock'',  just as the phase shift acquired by a system in the electromagnetic potential. Therefore the notion of proper time is not probed in such experiments. 

\subsection*{Massive quantum ``clock'' in an external gravitational field}\label{sec:hamilt}
In the next paragraphs we present how the above idea can be realized when the ``clock'' degrees of freedom are implemented in internal states of a massive particle (neglecting the finite-size effects).  Let $\h$ be the Hamiltonian that describes the internal evolution. In the rest reference frame, the time coordinate corresponds to the proper time $\tau$ and the evolution of the internal states is given by $i\hbar \frac{\partial}{\partial \tau} = \h$. Changing coordinates to the laboratory frame, the evolution is given by $i \hbar\frac{\partial}{\partial t}  = \dot{\tau} \h $, where $\dot{\tau}=\frac{d \tau}{d t}$  describes how fast the proper time flows with respect to the coordinate time. For a general metric $g_{\mu\nu}$, it is given by $\dot{\tau}=\sqrt{-g_{\mu\nu}\dot{x}^{\mu}\dot{x}^{\nu}}$, where we use the signature $(- + + +)$ and summation over repeated indices is understood. The energy-momentum tensor of a massive particle described by the action $S$ can be defined as the functional derivative of $S$ with respect to the metric, i.e.\ $T^{\mu\nu}:=\frac{\delta S}{\delta g_{\mu\nu}}$ (see e.g\ Ref.\ \cite{weinberg}). Since the particle's energy $E$ is defined as the $T_{00}$ component, it reads $E=g_{0\mu} g_{0\nu}T^{\mu\nu}$. In the case of a free evolution in a space-time with a stationary metric  (in coordinates such that $g_{0j}=0$ for $j=1,2,3$) we have
\begin{equation}
\label{energy}
E =  mc^2 \frac{-g_{00}}{\sqrt{-g_{\mu\nu}\dot{x}^{\mu}\dot{x}^{\nu}}},
\end{equation}
where $m$ is the mass of the particle. 
Space-time geometry in the vicinity of Earth can be described by the  Schwarzschild metric. In isotropic coordinates $(x, \theta, \vartheta)$ and with $d\Omega^2\equiv d\theta^2+\sin^2\theta\,d\vartheta^2$ it takes the form  \cite{weinberg}  $c^2 {d \tau}^{2} = \frac{(1+\frac{\phi(x)}{2c^2})^{2}}{(1-\frac{\phi(x)}{2c^2})^{2}} \, c^2 {d t}^2 - \left(1-\frac{\phi(x)}{2c^2}\right)^{4}\left(dx^2 + x^2 d\Omega^2\right)$, where $\phi(x) = - \frac{GM}{x}$ is the Earth's gravitational potential ($G$ denotes the gravitational constant and $M$ is the mass of Earth).
We consider the limit of a weak field and of slowly moving particles. In the final result we therefore keep up to quadratic terms in the kinetic and potential energy. In this approximation the metric components read  \cite{weinberg} $g_{00}\simeq-\left(1+2\frac{\phi(x)}{c^2}+ 2\frac{\phi(x)^2}{c^4} \right)$, $g_{ij}\simeq \frac{1}{c^2} \delta_{ij}\left(1-2\frac{\phi(x)}{c^2}\right)$, so that $\dot{\tau} \simeq \sqrt{1 + 2\frac{\phi(x)}{c^2} + 2\frac{\phi(x)^2}{c^4} - \left(\frac{\dot x}{c}\right)^2\left(1-2\frac{\phi(x)}{c^2} \right)}$. The total Hamiltonian in the laboratory frame is given by $H_{Lab} = H_0 + \dot{\tau} \h$, where the operator $H_0$ describes the dynamics of the external degrees of freedom of the particle and is obtained by canonically quantizing the energy \eqref{energy}, i.e. the particle's coordinate $x$ and kinematic momentum  $p=m\dot{x}$ become operators satisfying the canonical commutation relation ($[x,p]=i \hbar$). Thus, approximating up to the second order also in the internal energy, $H_{Lab}$ reads
\begin{equation}
\label{lab}
\hlab  \simeq mc^2 + \h + E_{k}^{GR}+\frac{\phi(x)}{c^2} \left(mc^2 + \h + E_{corr}^{GR} \right),
\end{equation}
where $E_{k}^{GR} = \frac{p^2}{2m}\left(1+3\left(\frac{p}{2mc}\right)^2-\frac{1}{mc^2} \h\right)$ and $E_{corr}^{GR} =\frac{1}{2}m\phi(x)-3\frac{p^2}{2m} $. We consider a semiclassical approximation of the particle's motion in the interferometer. Therefore, all terms in $\hlab$,  apart from the internal Hamiltonian $\h$, appear as purely numerical functions defined along the fixed trajectories.

In a setup as in Fig. 1, the particle follows in superposition two fixed non-geodesic paths $\gamma_1$, $\gamma_2$ in the homogeneous gravitational field. The acceleration and deceleration, which the particle undergoes in the $x$ direction, is assumed to be the same for both trajectories, as well as the constant velocity along the $y$ axis. This assures that the trajectories have different proper length but there will be no time dilation between the paths stemming from special relativistic effects. The particle inside the interferometer will thus be described by  the superposition  $|\Psi_{MZ}\ket  =\frac{1}{\sqrt{2}} \left(i|\Psi_{1}\ket + e^{i \varphi} |\Psi_{2}\ket\right) $, where the states $|\Psi_i\ket$ associated with the two paths $\gamma_i$ are given by applying the Hamiltonian \eqref{lab} to the initial state, which we denote by $|x^{in}\ket|\tau^{in}\ket$. Up to an overall phase these states read
\begin{equation}
\label{psi_i}
|\Psi_{i}\ket=e^{-\frac{i}{\hbar}\int_{\gamma_i}dt\; \frac{\phi(x)}{c^2} \left(mc^2 + \h + E_{corr}^{GR} \right)}|x^{in}\ket|\tau^{in}\ket.
\end{equation}
For a small size of the interferometer the central gravitational potential $\phi(x)$ can be approximated to linear terms in the distance $\Delta h$ between the paths:
\begin{equation}
\label{potential}
\phi(R+ \Delta h) = \phi(R)+g \Delta h+ \mathcal{O}({\Delta h}^2)\;,
\end{equation}
where $g=\frac{G M}{R^2}$ denotes the value of the Earth's gravitational acceleration in the origin of the laboratory frame, which is at distance $R$ from the centre of Earth.

For a particle having two internal states $|0\ket$, $|1\ket$ with corresponding energies $E_0$, $E_1$, the rest frame Hamiltonian of the internal degrees of freedom  can be written as 
\begin{equation}
\label{clock}
\h = E_0|0\ket\bra 0| + E_1|1\ket\bra 1|
\end{equation}
and if we choose the initial state of this internal degrees to be 
\begin{equation}
\label{tau0}
|\tau^{in}\ket=\frac{1}{\sqrt{2}}(|0\ket+|1\ket)
\end{equation}
the detection probabilities read
\begin{equation}
\label{eqn:probab}
P_{ \pm}(\varphi, m, \Delta E, \Delta V, \Delta T )=
\frac{1}{2} \pm \frac{1}{2}\cos\left(\frac{\Delta E  \Delta V \Delta T}{2 \hbar c^2}\right)\cos\left( (mc^2+\bra\h\ket + \bar{E}^{GR}_{corr}) \frac{ \Delta V \Delta T}{\hbar c^2}+\varphi\right),
\end{equation}
where $\Delta T$  is the time (as measured in the laboratory frame) for which the particle travels in the interferometer in a superposition of  two trajectories at constant heights, $\Delta V:=g \Delta h$ is the difference in the gravitational potential between the paths, $\bar{E}^{GR}_{corr}$ represents the
the corrections $E^{GR}_{corr}$ from Eq.\ \eqref{lab} averaged over the two trajectories and $\Delta E  := E_1-E_0$.  The expectation value $\bra \h \ket$ is taken with respect to the state \eqref{tau0}.
The corresponding visibility \eqref{visib_tau} is
\begin{equation}
\label{visib_wwi}
\mathcal V=\left| \cos \left( \frac{\Delta E  \Delta V \Delta T}{2 \hbar c^2} \right) \right|.
\end{equation}

\begin{figure}[h]
\vspace{-5pt}
\begin{center}
\includegraphics[width=12cm]{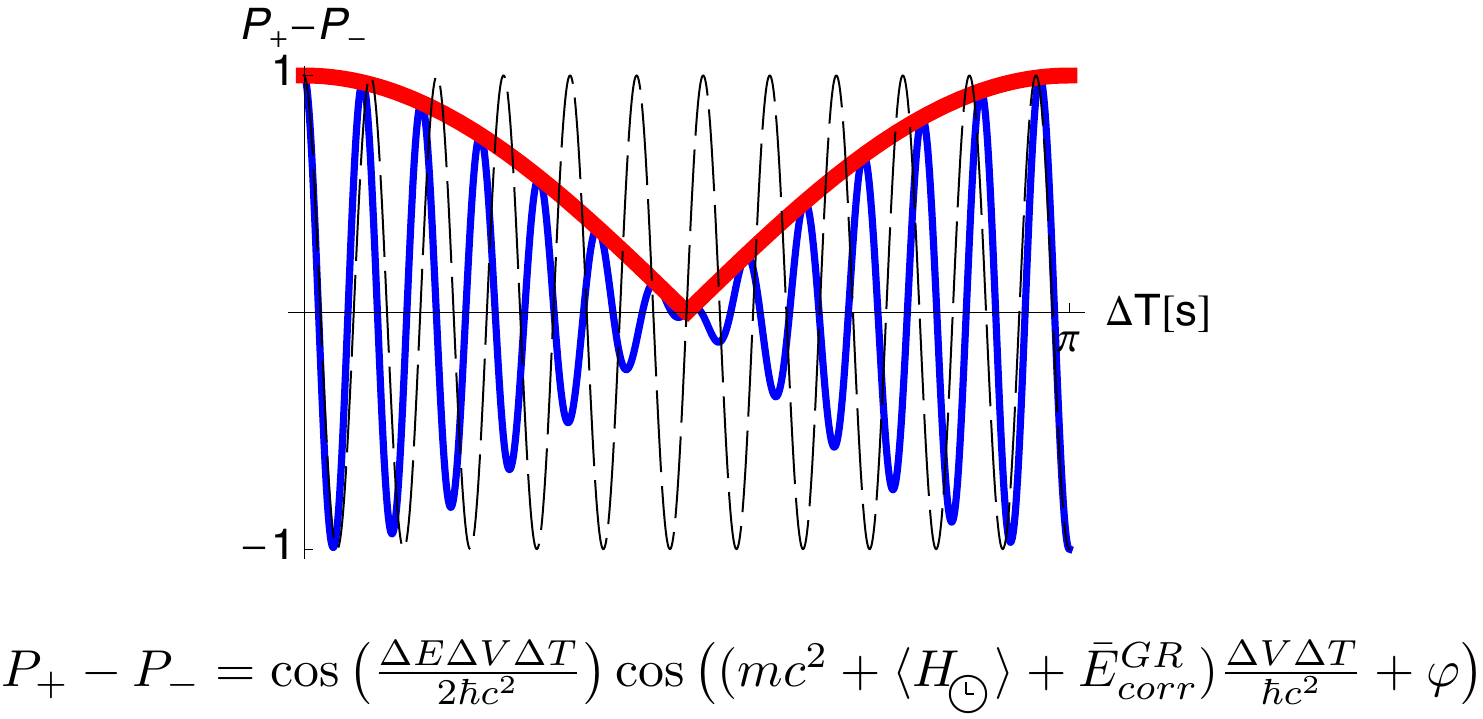}
\vspace{-5pt}
\caption{ The plot of the difference between the probabilities $P_{\pm}(\varphi, m, \Delta E, \Delta V, \Delta T )$, Eq.\ \eqref{eqn:probab},  to find the particle in the output path of the Mach-Zehnder interferometer as a function of  the time $\Delta T$  for which the particle travels in a superposition of two trajectories at constant heights (this corresponds to changing the length of the interferometric arms). The term proportional to the particle's  mass is the phase originating from the Newtonian potential energy $m \Delta V$. General relativistic corrections stemming from external degrees of freedom are given by $\bar{E}^{GR}_{corr}$, see e.g.\  Ref.\ \cite{wajima}. Without the ``clock'' degrees of freedom only these terms are present in the result  (dashed, black line in the plot). In the situation with the ``clock'' (blue line) we expect  two new effects: the change of the interferometric visibility given by the absolute value of the first cosine (thick red line) and an additional phase shift proportional to the average internal energy of the ``clock''. The values for the energy gap $\Delta E$ and the gravitational potential difference $\Delta V$ between the interferometric paths are chosen such that $\frac{\Delta E \Delta V}{2 \hbar c^2}=1$Hz. While the phase shift alone can always be understood as an Aharonov-Bohm phase of an effective potential, the notion of general relativistic proper time is necessary to explain the decrease of the visibility.\hspace*{\fill}}
\label{fig:probab}
\end{center}
\end{figure}

The introduction of the ``clock" degrees of freedom results in two new quantum effects that cannot be explained without including general relativity: the change of the interferometric visibility and the extra phase shift proportional to the average internal energy (Fig.\ \ref{fig:probab}  and  Eq.\ \eqref{eqn:probab}). The drop in the
visibility is a consequence of a direct coupling of the particle's
internal degrees of freedom to the potential in the effective Hamiltonian
\eqref{lab}. Such a coupling is never found in Newtonian gravity and it is
the mathematical expression of the prediction that  the ``clock''  ticks
at different rates when placed in different gravitational potentials.
This coupling can directly be obtained from the Einstein equivalence
principle. Recall that the latter postulates that accelerated reference
frames are physically equivalent to those in the gravitational field of
massive objects. When applied within special relativity this exactly results in the  
prediction that initially synchronized clocks subject to different gravitational potentials will show different times when brought together. The proposed experiment probes the presence of such a gravitational time dilation effect for a quantum system - it directly shows whether the
``clock'' would tick at different rates when taken along the two possible trajectories in the interferometer.
On the other hand, to obtain the correct phase shift it is sufficient to consider a
semiclassical coupling of the average total energy of the system to the
gravitational potential. With such a coupling the time displayed by the ``clock'' used in the
experiment will not depend on the path taken. This means that
a gravitationally induced phase shift can probe general relativistic
corrections to the Newtonian gravitational potential but is always
consistent with having an operationally well defined notion of global
time, i.e. with a flat space-time. 

The effect described in our work  follows directly from the Einstein equivalence principle, which is itself crucial for the formulation of general relativity as a metric theory \cite{clifford}. Thus, the drop in the fringe contrast is not only genuinely quantum mechanical but also a genuine general relativistic effect that in particular unambiguously probes the general relativistic notion of proper time.

\subsection*{General ``clocks'' and  gravitational fields}\label{sec:intefer}
Let us call $t_\perp$  the orthogonalization time of a quantum system, i.e. the minimal time needed for a quantum state to evolve under a given Hamiltonian into an orthogonal one \cite{mandelstam, fleming}.  For the initial state \eqref{tau0} subject to the rest frame Hamiltonian $\h$ given by Eq.\ \eqref{clock} we obtain
\begin{equation}
\label{omega}
t_\perp=\frac{\pi \hbar}{\Delta E}. 
\end{equation}
A system with finite $t_\perp$ can be seen as a clock that ticks at a rate proportional to $t_\perp^{-1}$. Thus, the orthogonalization time gives also the precision of a considered ``clock''. From the expression for $\dot\tau$ in the approximation \eqref{potential} it follows that the total time dilation $\Delta \tau$ between the trajectories is 
\begin{equation}
\label{deltatau}
\Delta \tau=\frac{\Delta V \Delta T}{c^2}.
\end{equation}
We can therefore phrase the interferometric visibility $\mathcal V$ solely in terms of $t_{\perp}$ and $\Delta \tau$:
\begin{equation}
\label{visib2}
\mathcal V=\left| \cos \left( \frac{\Delta \tau}{t_{\perp}}\frac{\pi}{2} \right) \right|.
\end{equation}
The total time dilation $\Delta \tau$ is a parameter capturing the relevant information about the paths and $t_{\perp}$ grasps pertinent features of the ``clock''. It is only their ratio which matters for the fringe visibility. Equation \eqref{visib2} is a generalization of the result \eqref{visib_wwi} to the case of an arbitrary initial state, ``clock'' Hamiltonian and a non-homogeneous gravitational field: whenever the time dilation $\Delta \tau$ between the two trajectories through the Mach-Zehnder interferometer is equal to the orthogonalization time $t_\perp$ of the quantum mechanical system that is sent through the setup, the physically accessible proper time difference will result in the full loss of fringe contrast. There are several bounds on the orthogonalization time based on energy distribution moments \cite{mandelstam,margolus,zielinski}.  Such bounds can through Eq.\ \eqref{visib2} give some estimates on the gravity induced decoherence rates in more general situations. As an example, for mixed states one generally has \cite{zielinski}:  $\frac{1}{t_\perp} \leq \frac{2^{\frac{1}{\alpha}}}{\pi \hbar}\bra (H-E_{gr})^\alpha \ket ^\frac{1}{\alpha}$, $\alpha >0$ (provided the initial state is in the domain of $(H-E_{gr})^{\alpha}$) where $H$ denotes the internal Hamiltonian and $E_{gr}$ the energy of the corresponding ground state. 

\section*{Discussion}\label{sec:discussion}
Current approaches to test general relativistic effects in quantum mechanics  mainly focus on high precision measurements of  the phase induced by the gravitational potential.  Although such experiments would probe the potential and thus could verify non-Newtonian corrections in the Hamiltonian, they would not constitute an unambiguous proof of the gravitational time dilation, since they are also explainable without this concept by the Aharonov-Bohm effect: a trajectory dependent phase acquired by a particle moving in a flat space-time in a presence of a position dependent potential. 

In our proposed experiment the effects arising from the general relativistic  proper time can be separated and probed independently from the Aharonov-Bohm type of effects.   
Unlike the phase shift, which occurs independently of whether the interfering particle can be treated as a ``clock'', the change of the interferometric visibility, Eq. \eqref{visib_wwi}, is a quantum effect that arises if and only if general relativistic proper time has a well defined operational meaning. Indeed, if one prepares the initial state $|\tau^{in}\ket$ as an eigenstate of the internal energy Hamiltonian $\h$, only the phase of such a state would change during the time evolution and according to Eq. \eqref{visib2} interferometric visibility would be maximal. This ``clock'' would not ``tick''  (it has orthogonalization time $t_{\perp}=\infty$) so the concept of proper time would have no operational meaning in this case. Moreover, reasoning that any (even just an abstract) frequency  which can be ascribed to the particle allows considering proper time as a physical quantity would imply that interference should always be lost, as the which-path information is stored ``somewhere''.  This once again shows that in quantum mechanics it makes no sense to speak about quantities without specifying how they are measured. 


The interferometric experiment proposed in this work can also be used to test whether proper time is a new quantum degree of freedom. This idea was discussed in the context of e.g.\ the equivalence principle in Refs.\ \cite{greenberger1, greenberger2} and a mass - proper time uncertainty relation \cite{matsumoto}. The equations of motion for proper time treated dynamically, as put forward in Refs.\ \cite{greenberger1, greenberger2, matsumoto}, are in agreement with general relativity. Therefore, the predictions of Eq.\ \eqref{visib_tau} would also be valid, if the states $|\tau_i\ket$, introduced in Eq.\ \eqref{intermed_tau}, stand for this new degree of freedom. Already performed experiments, like in Refs.\ \cite{cow, muller}, which measured a gravitational phase shift, immediately rule out the possibility that the state of proper time was sharply defined in those tests, in the sense of $\bra \tau_1|\tau_2\ket = \delta(\tau_1 - \tau_2)$. However, such experiments can put a finite bound on the possible uncertainty in the state of proper time. The phase shift measured in those experiments can be phrased in terms of the  difference in the proper time  $\Delta \tau$ between the paths. Denote by $\Delta \mathcal V$ the experimental error with which the visibility of the interference pattern was measured in those tests.  As a result, a Gaussian state of the proper time degree of freedom of  width  $\sigma_{\tau}$  such that $\sigma_\tau > \frac{|\Delta \tau|}{\sqrt{-8 \ln(1-\Delta \mathcal V)}}$, is consistent with the experimental data.  An estimate of the proper time uncertainty can be based on the Heisenberg uncertainty principle for canonical variables and the equation of motion for the proper time. In such an analysis the rest mass $m$ can be considered as a canonically conjugated momentum to the proper time variable $\tau$, i.e.\ one assumes $[\tau, mc^2 ] = i\hbar$ \cite{greenberger1, greenberger2, matsumoto}. 
In table  \ref{table:timeket_predictions} we discuss what can be inferred about  proper time as a quantum degree of freedom from an experiment in which the measured visibility would be $\mathcal V_{m}$ and where $\mathcal V_{QM}$ is the visibility predicted by quantum mechanics as given by Eq.\ \eqref{visib_wwi}.

\captionsetup[table]{labelformat=default, labelsep=default, justification=centerlast, font=small}
\begin{table}[H]
\caption{Discussion of possible outcomes of the proposed interferometric experiment. The measured visibility $\mathcal V_{m}$ is compared with the quantum mechanical prediction $\mathcal V_{QM}$ given by \eqref{visib_wwi}. Depending on their relation different conclusions can be drawn regarding the possibility that proper time is a quantum degree of freedom (d.o.f.). Assuming that the distribution of the proper time d.o.f. is a Gaussian of the width $\sigma_\tau$, current interferometric experiments give bounds on possible $\sigma_\tau$ in terms of  the proper time difference $\Delta \tau$ between the paths and the experimental error  $\Delta \mathcal V$ of the visibility measurement. \hspace*{\fill}} 
\centering 
\begin{tabular}{c c c c c} 
\hline  
  \textbf{ experimental visibility  }                                      &      \textbf{possible  explanation}           &\textbf{current  experimental status}     \\ \hline
 $\mathcal {V}_{m}=0$                                                       &  proper time: quantum d.o.f.,                   &    disproved in   \\
                                                                                                   & sharply defined                                      &   e.g.\ Refs.\ \cite{cow, chu} \\ \hline
  $0<\mathcal {V}_{m}<\mathcal {V}_{QM}$		     &   proper time: quantum d.o.f                 &   consistent with current data  \\
                            								     &    with uncertainty $\sigma_\tau$        &  for $\sigma_\tau > \frac{|\Delta \tau|}{\sqrt{-8 \ln(1-\Delta \mathcal V)}}$  \\ \hline           
$\mathcal {V}_{m}=\mathcal {V}_{QM}$                         &  proper time: not a quantum d.o.f.                       &    consistent  with current data   \\
                                            						     &  or has a very broad uncertainty                     	       &         \\	 \hline
$\mathcal {V}_{m}>\mathcal {V}_{QM}$                          &  quantum interferometric complementarity              &   not  tested         \\
                                                                                                 &  does  not hold  when  general            &                                 \\
 										     &  relativistic effects become relevant                    &        \\
\end{tabular}  
\label{table:timeket_predictions} 
\end{table}

\subsection*{Conclusion}\label{sec:conclusion}
In conclusion, we predicted a quantum effect in interferometric experiments that for the first time 
allows probing general relativistic proper time in an unambiguous way. In the presence of a gravitational potential, we showed that  a loss in the interferometric visibility occurs if the time dilation is physically accessible from the state of the interfered particle. This requires that the particle is a ``clock'' measuring proper time along the trajectories, therefore revealing the which-way information. Our predictions can be experimentally verified by implementing the ``clock'' in some internal degrees of freedom of the particle, see Sec.\ Methods.
The proposed experiment can also lead to a conclusive test of theories in which proper time is treated as a quantum degree of freedom. 
As a final remark we note that decoherence due to the gravitational time dilation may have further importance in considering the quantum to classical transition and in attempts to observe collective quantum phenomena in extended, complex quantum systems since the orthogonalization time may become small enough in such situations to make the predicted decoherence effect prominent.

\section*{Methods}\label{sec:methods}
Here we briefly discuss various systems for the possible implementation of the interferometric setup. Interferometry with many different massive quantum systems has been achieved, e.g.\ with neutrons \cite{cow, rauch}, atoms \cite{muller, muller2}, electrons \cite{neder, ji}, and molecules \cite{arndt, gerlich}. In our framework, additional access to an internal degree of freedom is paramount, as to initialize the ``clock'' which measures the proper time along the interferometric path. Therefore, the experimental requirements are more challenging. To observe full loss of the interferometric visibility, the proper time difference in the two interferometric arms needs to be $ \Delta \tau = t_{\perp}$. For a two-level system the revival of the visibility due to the indistinguishability of the proper time in the two arms occurs when  $\Delta \tau =  2 \, t_{\perp}$.
 
The best current atomic clocks operate at optical frequencies $\omega$ around $10^{15}$ Hz. For such systems, we have $t_{\perp} = \frac{\pi}{\omega}$ and one would therefore require an atomic superposition with $\Delta h \Delta T \approx 10$ \ms in order to see full disappearance of the interferometric visibility. For example, the spatial separation would need to be of the order of 1 m, maintained for about 10 s. Achieving and maintaining such large superpositions of atoms still remains a challenge, but recent rapid experimental progress indicates that this interferometric setup could be conceivable in the near future. For neutrons, a separation of $\Delta h \sim 10^{-2}$ m with a coherence time of $t \sim 10^{-4}$ s has been achieved \cite{rauch}. To implement our ``clock'' in neutron interferometry one can use spin precession in a strong, homogeneous magnetic field. However, such a ``clock'' could reach frequencies up to $\omega \sim 10^{9}$ Hz (for a magnetic field strength of order of $10$ T \cite{miller}), which is still a few orders of magnitude  lower than necessary for  the observation of full decoherence due to a proper time difference. Improvements in the coherence time and the size of the interferometer would still be necessary. Other systems, such as molecules, could be used as well and table \ref{table:expparamters} summarizes the requirements for various setups (note again that the particles are assumed to travel at fixed height during the time $\Delta T$).

\captionsetup[table]{labelformat=default, labelsep=default, justification=centerlast, font=small}
\begin{table}[H]
\caption{Comparison of different systems for the experimental observation of  the reduced interferometric visibility. Several possible systems are compared on the basis of theoretically required and already experimentally achieved parameters, which are relevant for our proposed experiment. For a ``clock''  with a frequency  $\omega =  \frac{\Delta E}{\hbar}$, the required value of the parameter $\Delta h\Delta T$ ($\Delta h$ being the separation between the interferometers arms and $\Delta T$ the time for which the particle travels  in superposition at constant heights) for the full loss of the fringe visibility, see Eq.\ \eqref{visib_wwi}, is given in the rightmost column. In our estimations we assumed a constant gravitational acceleration $g = 10 \frac{m}{s^2}$. See section Methods for further discussion on possible experimental implementations.\hspace*{\fill}}
\centering 
\begin{tabular}{c c c c c} 
\hline
  \textbf{system}   &   \textbf{ ``clock''  }                       & $\mathbf{\omega}$ {\bf[Hz]} & $\mathbf{\Delta h\Delta T}$ {\bf[m$\cdot$s]} & $\mathbf{ \Delta h\Delta T}$ {\bf[m$\cdot$s]}  \\
                      &                                   &                            & {\bf achieved}                  &  {\bf required}     \\\hline
atoms           & hyperfine states      & $10^{15}$        & $10^{-5}$                         & $10$               \\ 
electrons     &  spin precession    &  $10^{13} $       & $10^{-6}$                          & $10^{3}$      \\
molecules   & vibrational modes  & $10^{12}$         & $10^{-8}$                          & $10^{4}$        \\
neutrons      & spin precession     & $10^{10}$         & $10^{-6}$                          & $10^{6}$        \\
\end{tabular}
\label{table:expparamters} 
\end{table}


The effect we predict can be measured even without achieving full orthogonalization of the ``clocks''. Note that even for  $\Delta \tau   \ll   t_{\perp}$ the small reduction of visibility can already be sufficient to prove the accessibility of which-path information due to the proper time difference. With current parameters in atom interferometry, an accuracy of the measurement of the visibility of $\Delta \mathcal V = 10^{-6}$ would have to be achieved for the experimental confirmation of our predictions. A very good precision measurement of the interferometric visibility and a precise knowledge about other decoherence effects would therefore make the requirements for the other parameters less stringent.

\begin{acknowledgments}
The authors thank M.\ Arndt, B.\ Dakic, S.\ Gerlich, D.\ M.\ Greenberger, H. MŸller, S.\ Nimmrichter, A.\ Peters, and P. Wolf for insightful discussions. The research was funded by the Austrian Science Fund (FWF) projects: W1210,  P19570-N16 and SFB-FOQUS, the Foundational Questions Institute (FQXi) and the European Commission Project Q-ESSENCE (No. 248095).  F.C., I.P., and M.Z.\ are members of the FWF Doctoral Program CoQuS.
\end{acknowledgments}

\begin{picture}(60,50)

\end{picture}


\begin{thebibliography}{99}

\bibitem{hafele}
Hafele, J.\ C.\  and  Keating, R.\ E.\
Around-the-world atomic clocks: predicted relativistic time gains.
\textit{Science} \textbf{177,}  166-168 (1972).
\bibitem{chiao} 
 Chiao, R.\ Y.\,  Minter, S.\ J., Wegter-McNelly,  K., and Martinez, L.\ A. 
Quantum incompressibility of a falling Rydberg atom, and a gravitationally-induced charge separation effect in superconducting systems.
\textit{Found. Phys.} 1-19 DOI: 10.1007/s10701-010-9531-2 (2011).
 
\bibitem{wajima}
Wajima, S., Kasai, M. and Futamase, T.
Post-Newtonian effects of gravity on quantum interferometry.
\textit{Phys. Rev. D} {\bf55,} 1964-1970 (1997).
 

\bibitem{dimopoulos}
Dimopoulos,S., Graham, P.\ W., Hogan, J.\ M.\ and Kasevich, M.\ A.\ 
General relativistic effects in atom interferometry.
\textit{Phys. Rev. D} {\bf78,} 042003 (2008). 
 
 \bibitem{aharonov}
Aharonov, Y.\ and Bohm, D.
Significance of electromagnetic potentials in the quantum theory.
\textit{Phys. Rev.} {\bf115,} 485-491 (1959). 

\bibitem{morgan}
 Ho, V.\ B.\ and Morgan, M.\ J.
 An experiment to test the gravitational Aharonov-Bohm effect.
\textit{ Aust. J. Phys.} {\bf 47,} 245-253 (1994).
 
  
\bibitem{cow}
Colella, R., Overhauser, A.\ W.\ and Werner, S.\ A.
Observation of gravitationally induced quantum interference.
\textit{Phys. Rev. Lett.} {\bf34,} 1472-1474 (1975).

\bibitem{rauch}
Zawisky, M., Baron, M., Loidl, R.\ and Rauch, H.
Testing the world's largest monolithic perfect crystal neutron interferometer.
\textit{Nucl.\ Instrum.\ Methods Phys.\ Res.\ A} {\bf481,} 406-413 (2002).

\bibitem{chu}
Peters, A.,  Chung, K.\ Y.\ and Chu, S.
Measurement of gravitational acceleration by dropping atoms.
\textit{Nature} {\bf400,} 849-852 (1999).


\bibitem{wooters}
Wootters, W.\ K.\ and Zurek, W.\ H.
Complementarity in the double-slit experiment: Quantum nonseparability and a quantitative statement of Bohr's principle.
\textit{Phys. Rev. D} {\bf 19,} 473-484 (1979).

\bibitem{green_yasin}
Greenberger, D.\ M.\ and Yasin, A. 
Simultaneous wave and particle knowledge in a neutron interferometer.
\textit{Phys. Lett. A} {\bf 128,} 391-394 (1988).

\bibitem{englert}
Englert, B.\-G. 
Fringe visibility and which-way information: An inequality.
\textit{Phys. Rev. Lett.} {\bf 77,} 2154-2157  (1996).


\bibitem{greenberger1}
Greenberger, D.\ M.
Theory of particles with variable mass. I. Formalism.
\textit{J. Math. Physics} {\bf11,}  2329 (1970),

\bibitem{greenberger2}
Greenberger, D.\ M.
Theory of particles with variable mass. II. Some physical consequences.
\textit{J. Math. Physics} {\bf11,}  2341 (1970).

\bibitem{matsumoto}
Kudaka, S.\ and Matsumoto, S.
Uncertainty principle for proper time and mass.
\textit{J. Math. Phys.} {\bf40,} 1237  (1999).

\bibitem{muller} 
M\"{u}ller, H., Peters, A.\ and Chu, S. 
A precision measurement of the gravitational redshift by the interference of matter waves.
\textit{Nature} {\bf 463,} 926-929 (2010).

\bibitem{wolf_rep}
Wolf, P.\ et  al.
Atom gravimeters and gravitational redshift.
\textit{Nature} {\bf 467,} E1 (2010).

\bibitem{sinha}
Sinha, S.\ and Samuel, J. 
 Atom interferometers and the gravitational redshift.
\textit{Class. Quantum Grav.} {\bf28,} 145018 (2011).
 
 \bibitem{giulini}
 Giulini, D.
Equivalence principle, quantum mechanics, and atom-interferometric tests.
arXiv:1105.0749v1 (2011).

\bibitem{wolf}
Wolf, P.\ et  al.
Does an atom interferometer test the gravitational redshift at the Compton frequency?
\textit{Class. Quantum Grav.} {\bf28,} 145017 (2011).

\bibitem{weinberg}
Weinberg, S.
\textit{Gravitation and Cosmology}
John Wiley \verb &  Sons, 1972, New York.

\bibitem{clifford}
Will, C.\ M.\ 
\textit{Theory and Experiment in Gravitational Physics}
Cambridge University Press, 1993, Cambridge.

\bibitem{mandelstam}
Mandelstam, L.\ and Tamm, I.
The uncertainty relation between energy and time in non-relativistic quantum mechanics.
 \textit{Journ. Phys.} (USSR) {\bf9,} 249 (1945).
 
 \bibitem{fleming}
Fleming,  G.\ N. 
 A unitarity bound on the evolution of nonstationary states.
\textit{Nuovo Cim. A} {\bf16,} 232-240 (1973).
 
 \bibitem{margolus}
Margolus, N.\ and Levitin, L.\ B.
The maximum speed of dynamical evolution.
\textit{ Physica D} {\bf120,} 188-195 (1998).


\bibitem{zielinski}
Zielinski, B.\ and Zych, M. 
Generalization of the Margolus-Levitin bound.
\textit{Phys. Rev. A} {\bf74,} 034301 (2006). 

\bibitem{muller2}
M\"{u}ller, H., Chiow, S.,  Herrmann, S.\ and Chu, S.
Atom interferometers with scalable enclosed area.
\textit{Phys. Rev. Lett.} {\bf102,} 240403 (2009).


\bibitem{neder}
Neder, I., Heiblum, M., Mahalu, D.\ and Umansky, V.
Entanglement, dephasing, and phase recovery via cross-correlation measurements of electrons.
\textit{Phys.\ Rev.\ Lett. } {\bf98,} 036803 (2007). 

\bibitem{ji}
Ji, Y.\ et al.
An electronic MachÐZehnder interferometer.
\textit{Nature} {\bf422,} 415-418 (2003).

\bibitem{arndt}
Arndt, M.\ et al.
Wave-particle duality of C 60 molecules.
\textit{Nature} {\bf401,} 680-682 (1999).

\bibitem{gerlich}
Gerlich, S.\ et al.
Quantum interference of large organic molecules.
\textit{Nat.\  Commun.}  {\bf2}:263 doi: 10.1038/ncomms1263  (2011).


\bibitem{miller}
 Miller, J.\ R.
 The NHMFL 45-T hybrid magnet system: past, present, and future.
 \textit{IEEE Trans. Appl. Supercond.,} {\bf13,} 1385-1390  (2003).

\end{thebibliography}
\end{document}